\newif\ifpreprint
\preprintfalse
\preprinttrue
\documentclass{cjour}
\usepackage{epsf}

\newcommand{\la}        {\mathrel{\hbox{\rlap{\hbox{\lower4pt\hbox{$\sim$}}}\hbox{$<$}}}}
\newcommand{\ga}        {\mathrel{\hbox{\rlap{\hbox{\lower4pt\hbox{$\sim$}}}\hbox{$>$}}}}
\newcommand{\B}[1]      {\mbox{\boldmath$#1$}}
\newcommand{\Bs}[1]     {\mbox{\boldmath$\scriptstyle#1$}}
\newcommand{\rA}        {{\rm A}}
\newcommand{\rB}        {{\rm B}}
\newcommand{\bM}[1]     {\B{\sf M}^{#1}}
\newcommand{\sM}[1]     {{\sf M}^{#1}}
\newcommand{\tM}[1]     {\tilde{\sf M}{}^{#1}}
\newcommand{\bC}[1]     {\B{\sf C}^{#1}}
\newcommand{\sC}[1]     {{\sf C}^{#1}}

\newcommand{\btC}[1]    {\tilde{\B{\sf C}}{}^{#1}}
\newcommand{\bF}        {\B{F}}
\newcommand{\bx}        {\B{x}}
\newcommand{\by}        {\B{y}}
\newcommand{\bz}        {\B{z}}
\newcommand{\bR}        {\B{R}}
\newcommand{\eqb}[1]    {(\ref{#1})}
\newcommand{\eqn}[1]    {equation \eqb{#1}}
\newcommand{\Eqn}[1]    {Equation \eqb{#1}}
\newcommand{\eqns}[1]   {equations \eqb{#1}}

\newcommand{\Sec}[1]    {\S\ref{sec:#1}}
\newcommand{\fig}[1]    {Fig.~\ref{fig:#1}}
\newcommand{\Fig}[1]    {Figure~\ref{fig:#1}}
\newcommand{\Tab}[1]    {Table~\ref{tab:#1}}
\newcommand{\half}      {{\textstyle{1\over2}}}

\begin{document}


\title  {A Hierarchical ${\cal O}(N)$ Force Calculation Algorithm}
\author {Walter Dehnen}
\affil  {Max-Planck-Institut f\"ur Astronomie, 
         K\"onigstuhl 17, D-69117 Heidelberg, Germany}
\authorrunninghead{W.~Dehnen}
\titlerunninghead{A Hierarchical ${\cal O}(N)$ Force Calculation Algorithm}
\email  {dehnen@mpia.de}
\abstract{A novel code for the approximate computation of long-range forces
  between $N$ mutually interacting bodies is presented. The code is based on a
  hierarchical tree of cubic cells and features mutual cell-cell interactions
  which are calculated via a Cartesian Taylor expansion in a symmetric way, such
  that total momentum is conserved. The code benefits from an improved and
  simple multipole acceptance criterion that reduces the force error and the
  computational effort. For $N\ga10^4$, the computational costs are found
  empirically to rise sub-linear with $N$. For applications in stellar dynamics,
  this is the first competitive code with complexity ${\cal O}(N)$, it is faster
  than the standard tree code by a factor ten or more.}
\section{Introduction}\label{sec:intro}
\ifpreprint\nobreak\noindent\fi
In $N$-body simulations of stellar dynamics (or any other dynamics incorporating
long-range forces), the computation, at every time step, of the gravitational
forces between $N$ mutually interacting bodies dominates the operational effort.
In many situations, such as studies of collisionless stellar systems, the error
of these simulations is dominated by the noise in the distribution of the bodies,
whose number $N$ is just a numerical parameter. Therefore, instead of computing
the forces exactly by direct summation over all pairs of bodies, one may use
approximate but much faster methods, allowing substantially larger $N$ and hence
significantly reduced noise.

In stellar dynamics, one of the most commonly used approximate methods is the
Barnes \& Hut tree code \cite{BH} and its clones. With these methods, one first
sorts the bodies into a hierarchical tree of cubic cells and pre-computes
multipole moments of each cell. Next, the force at any body's position and
generated by the contents of some cell is calculated by a multipole expansion if
the cell is well-separated from the body; otherwise the forces generated by the
cell's child nodes are taken. This technique reduces the number of interactions
per body to ${\cal O}(\log N)$ and hence results in an overall complexity of
${\cal O}(N\log N)$.

Another technique used frequently, for instance in molecular dynamics, is
Greengard \& Rokhlin's \cite{FMMa,FMMa1} fast multipole method (FMM) and its
variants.  Traditionally, these methods first sort the bodies into a hierarchy
of nested grids, pre-compute multipole moments of each cell, and then compute
the forces between grid cells by a multipole expansion, usually in spherical
harmonics.  That is, cells are not only sources but also sinks, which formally
reduces the complexity to ${\cal O}(N)$, although the author is not aware of an
empirical demonstration in three dimensions\footnote{One of the most recent
  members of the FMM family, presented by Cheng, Greengard \& Rokhlin
  \cite{FMMb}, does still not show clear ${\cal O}(N)$ behavior at $N=10^6$
  (see, e.g., table I of their paper).}.

Currently, no useful implementation of FMM for application in stellar dynamics
exists. In fact, it has been shown \cite{CM} that, for this purpose, the FMM in
its traditional form cannot compete with the tree code. The reason, presumably,
is two-fold: first, because stellar systems are very inhomogeneous, non-adaptive
methods are less useful; second FMM codes are traditionally designed for high
accuracy, whereas in collisionless stellar dynamics a relative force error of
few $10^{-3}$ is often sufficient.

Here, we describe in detail a new code, designed for application in the
low-accuracy regime, that combines the tree code and FMM whereby taking the
better of each. In order to be fully adaptive, we use a hierarchical tree of
cubic cells. The force is calculated employing {\em mutual\/} cell-cell
interactions, in which both cells are source and sink simultaneously. Whether a
given cell-cell interaction can be executed or must be split, is decided using
an improved multipole-acceptance criterion (MAC). The new code is a further
development of the code presented in \cite{Da}, which in turn may be considered
an improvement of a code given in \cite{WS}. It is substantially faster than the
tree code and empirically shows a complexity of ${\cal O}(N)$ or even less.

The paper is organized as follows. In \Sec{grav}, the numerical concepts are
presented; \Sec{alg} describes the algorithm; in \Sec{assess} the force errors
are empirically assessed for some typical stellar dynamical test cases;
\Sec{perform} presents empirical measurements of the complexity and performance,
also in comparison to other methods; finally, \Sec{summ} sums up and concludes.

\section{Approximating Gravity}\label{sec:grav}
\ifpreprint\nobreak\noindent\fi
The goal is to approximately compute the gravitational potential $\Phi$ and its
derivative, the acceleration, at all body positions $\bx_i$ and generated by all
other $N-1$ bodies
\begin{equation} \label{potential}
\Phi(\bx_i) =-\sum_{j\neq
  i}\mu_j\,g(\bx_i-\bx_j),
\end{equation}
where $\mu_i$ is the weight of the $i$th body.
\begin{figure}
  \centerline{\epsfxsize=75mm \epsfbox[51 155 585 460]{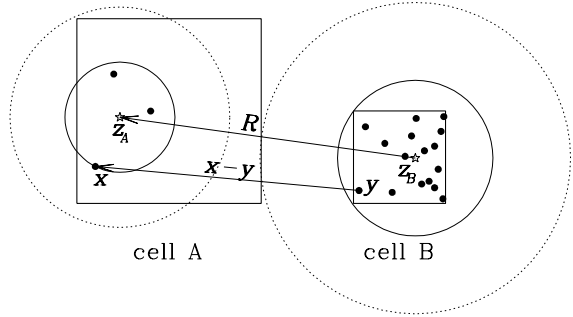}}
  \caption{Two interacting cells ({\it boxes}). Bodies are shown as solid
    dots. The stars indicate the positions of the centers of mass, the solid and
    dotted circles around of which have radii $r_{\rm max}$ and $r_{\rm crit}$,
    respectively, for $\theta=0.5$ \label{fig:cells}}
\end{figure}
Consider two cells A and B, see Figure~\ref{fig:cells}, with centers of mass
$\bz_\rA$ and $\bz_\rB$, respectively. The Greens function describing the mutual
interaction between a body at position $\bx$ in cell A and a body at position
$\by$ in cell B may be Taylor expanded about the separation
$\bR\equiv\bz_\rA-\bz_\rB$:
\begin{equation} \label{Taylor}
  g(\bx-\by) = \sum_{n=0}^p {1\over n!}\,(\bx-\by-\bR)^{(n)}
  \odot\B\nabla^{(n)}g(\bR)\;+\;{\cal R}_p(g),
\end{equation}
where $p$ is the order of the expansion, while ${\cal R}_p$ denotes the Taylor
series remainder (see also Appendix~A).  Here, we follow Warren \& Salmon
\cite{WS} by using the notational shorthand in which $\bx^{(n)}$ indicates the
$n$-fold outer product of the vector $\bx$ with itself, while $\odot$ denotes a
tensor inner product.  When inserting \eqb{Taylor} into \eqb{potential}, whereby
restricting the sum over $j$ to all bodies within cell B, we obtain for the
potential at every position $\bx$ in cell A and generated by all bodies in cell
B \cite{WS}
\begin{eqnarray} \label{local-expn}
  \Phi_{\rB\to\rA}(\bx) &=&-\sum_{m=0}^p {1\over m!}\,(\bx-\bz_\rA)^{(m)} 
                            \odot\bC{m,p}_{\rB\to\rA}
                            \;+\;{\cal R}_p(\Phi_{\rB\to\rA})
                            \\[1ex] \label{coeffs}
  \bC{m,p}_{\rB\to\rA}  &=& \sum_{n=0}^{p-m} {(-1)^n\over n!}\;\B\nabla^{(n+m)}
                            g(\bR) \odot \bM{n}_\rB,
                            \\[1ex] \label{multipoles}
  \bM{n}_\rB            &=& \sum_{\Bs{y}_i\in\rB} \mu_i\,(\by_i-\bz_\rB)^{(n)}.
\end{eqnarray}
The summation over $m$ in \eqn{local-expn} represents the evaluation of gravity,
represented by the field tensors $\bC{m,p}_{\rB\to\rA}$, at the evaluation point
$\bx$ within the sink cell A.  The computation of the field tensors via the
summation over $n$ in \eqn{coeffs} represents the interaction between sink cell
A and source cell B, represented by its multipole moments $\bM{n}_\rB$.

The symmetry between \bx\ and \by\ of the Taylor expansion \eqb{Taylor} has the
important consequence that, if \eqns{local-expn} and \eqb{coeffs} are used to
compute $\B\nabla\Phi_{\rB\to\rA}$ {\em and} $\B\nabla\Phi_{\rA\to\rB}$,
Newton's third law is satisfied by construction. For instance, the sum over all
forces of $N$ bodies vanishes within floating point accuracy.

Note that the highest-order multipole moments, $\bM{n=p}$, contribute only to
the coefficients $\bC{0,p}$, and hence affect only $\Phi$ but not
$\B\nabla\Phi$. Since, in stellar dynamics the acceleration rather than the
potential is to be computed, one may well ignore $\bM{p}$, reducing CPU-time and
memory requirements.

The formulae used in the standard tree code can be obtained by setting
$\bx=\bz_\rA$, corresponding to body sinks. In this case, potential and
acceleration are approximated by $-\bC{0,p}_{\rB\to\rA}$ and
$\bC{1,p}_{\rB\to\rA}$, respectively.

\subsection{Gravity Between Well-Separated Cells}\label{sec:hard}
\ifpreprint\nobreak\noindent\fi
In our implementation, we stick to a third order expansion ($p=3)$, whereby
ignoring octopoles $\bM3_\rB$ (see remark above). The dipole $\bM1_\rB$ vanishes
by construction and the Taylor-series coefficients for spherical Greens
functions read (with Einstein's sum convention)
\begin{equation}  \label{hard-coeffs}
  \begin{array}{lcl}
    \sC{0}_{\rB\to\rA}        &=& \sM{}_\rB \big[D^0
                                  + \half\tM2_{\rB ii}\,D^1
                                  + \half R_iR_j\tM2_{\rB ij}D^2
                                    \big],                           \\[1ex] 
    \sC{1}_{\rB\to\rA,\,i}    &=& \sM{}_\rB \big[R_i\big(D^1      
                                  + \half\tM2_{\rB jj}D^2
                                  + \half R_jR_k\tM2_{\rB jk}D^3\big)
                                  + R_j\tM2_{\rB ij}D^2
                                    \big],                           \\[1ex]
    \sC{2}_{\rB\to\rA,\,ij}   &=& \sM{}_\rB \big[
                                    \delta_{ij}\,D^1+ R_iR_j\,D^2
                                    \big],                           \\[1ex]
    \sC{3}_{\rB\to\rA,\,ijk}  &=& \sM{}_\rB \big[\big(
                                    \delta_{ij}R_k+\delta_{jk}R_i
                                   +\delta_{ki}R_j\big)D^2
                                   +R_iR_jR_kD^3\big],
  \end{array}
\end{equation}
where $\sM{}_\rB\equiv\sM0_\rB$ is the mass of cell B and
$\tM2_\rB\equiv\sM2_\rB /\sM0_\rB$ its {\em specific\/} quadrupole moment, while
\begin{equation} \label{Dn}
  D^m\equiv \left.
    \left({1\over r}{\partial\over\partial r}\right)^m
    g(r) \right|_{r=|\Bs{R}|}.
\end{equation}
In practice, we calculate the coefficients $\btC{m}_{\rB\to\rA}
\equiv\sM{}_\rA\bC{m}_{\rB\to\rA}$, because these obey the symmetry relations
$\btC{2}_{\rB\to\rA}= \btC{2}_{\rA\to\rB}$ and
$\btC{3}_{\rB\to\rA}=-\btC{3}_{\rA\to\rB}$ (for $p=3$), which arise from the
mutuality of gravity. Since we consider always both directions of any
interaction, exploiting these relations substantially reduces the operational
effort.
\subsection{Accumulating Taylor Coefficients}\label{sec:accum}
\ifpreprint\nobreak\noindent\fi
After, for each cell, the coefficients of all its interactions have been
accumulated as $\btC{m}_\rA = \sum_\rB \btC{m}_{\rB\to\rA}$, where the sum
includes all interaction partners B of cell A, we transform back to $\bC{m}_\rA
=\btC{m}_\rA /\sM{}_\rA$. Next, for each body, the Taylor series of all relevant
cells (those that contain the body), have to be accumulated by first translating
to a common expansion center and then adding coefficients.

In contrast to expansions in spherical harmonics, the translation of the Cartesian
expansion \eqb{local-expn} to a different center $\bz$ is straightforward. Let
$\bC{m,p}_0$ be the coefficients for expansion center $\bz_0$, then the
coefficients for expansion center $\bz_1$ are
\begin{equation} \label{translate}
  \bC{m,p}_1 = \sum_{n=0}^{p-m} {1\over n!}\,(\bz_0-\bz_1)^{(n)} \odot
  \bC{m+n,p}_0.
\end{equation}
\subsection{The Multipole Acceptance Criterion (MAC)}\label{sec:mac}
\ifpreprint\nobreak\noindent\fi
For the expansion \eqb{local-expn} to converge, we must have $|\bx-\by-\bR|
<|\bR|$ for all body-body interactions `caught' by a single cell-cell (or
cell-body) interaction (see Appendix~A). In order to ensure this, we first
obtain, for each node, an upper limit $r_{\rm max}$ for the distance of any body
within the node from the center of mass (bodies naturally have $r_{\rm max}=0$).
We take $r_{\rm max}$ to be either the distance from the cell's center of mass
to its most distant corner \cite{SW}, or \cite{Benz} 
\begin{equation}
  \max_{{\rm child\;nodes}\;i}\{r_{i,\rm max}+|\bz-\bz_i|\},
\end{equation}
whatever is smaller. Then $|\bx-\by-\bR|<\theta|\bR|\;\forall\;\bx \in\rA\;{\rm
  and}\;\by\in\rB$, i.e.\ the nodes are {\em well-separated\/}, if
\begin{equation} \label{well-separated}
  |\bR| > r_{\rm A,crit} + r_{\rm B,crit}
  \quad{\rm with}\quad
  r_{\rm crit} = r_{\rm max} / \theta.
\end{equation}
The {\em tolerance parameter\/} $\theta$ controls the accuracy of the
approximation: for Newtonian forces, the error made in $d$ dimensions by the
$p$th order expansion of the form \eqb{local-expn} is (\eqn{error:acc})
\begin{eqnarray} \label{error:rel}
  |{\cal R}_p(\B\nabla\Phi_{\rB\to\rA})| &\le&
  {(p+1)\theta^p\over(1-\theta)^2}\;{\sM{}_\rB\over R^2}
\\   \label{error:abs} &\propto&
  {\theta^{p+2}\over(1-\theta)^2}\;r_{\rm B,max}^{d-2}
  \propto
  {\theta^{p+2}\over(1-\theta)^2}\;{\sM{}_\rB}^{(d-2)/d}
\end{eqnarray}
where $\sM{}_\rB\propto r_{\rm B,max}^d$ has been assumed. Since $\sM{}_\rB/R^2$
is, to lowest order, the acceleration due to the interaction, \eqn{error:rel}
tells us that with constant $\theta$, which is standard tree-code practice, the
{\em relative\/} error introduced by every single interaction is approximately
constant. \Eqn{error:abs}, however, shows that for the most important case of
$d=3$ this practice results in larger {\em absolute\/} errors for interactions
with bigger and hence on average more massive cells.  This implies that these
interactions will dominate the {\em total\/} error of any body's acceleration.
It is, therfore, expedient to balance the absolute errors of the individual
interactions, which can be approximately achieved by using a mass-dependent
tolerance parameter as follows
\begin{equation} \label{open:mass}
        {\theta^{p+2}\over(1-\theta)^2} =
        {\theta_{\rm min}^{p+2}\over(1-\theta_{\rm min})^2}
        \left({\sM{}\over\sM{}_{\rm tot}}\right)^{(2-d)/d}
\end{equation}
where $\sM{}_{\rm tot}$ is the mass of the root cell, while $\theta_{\rm min}=
\theta(\sM{}_{\rm tot})$ is the new tolerance parameter. If $r_{\rm A,max}\sim
r_{\rm B,max}$ and $\sM{}_\rA\sim\sM{}_\rB$, this method results in
approximately constant absolute acceleration errors. Note that \eqn{open:mass}
results in a very weak decrease of $\theta$ with increasing mass:
$\theta\propto\sM{-1/15}$ for $d=3$, $p=3$ and $\theta\ll1$.

Instead of \eqn{error:abs}, one can obtain a stricter error limit incorporating
the first $p+1$ multipole moments (see Appendix~A), which may be turned into an
MAC \cite{WS}. However, our choice \eqb{open:mass} is (i) much simpler and (ii)
overcomes already the main disadvantage of $\theta=\,$const, the variations of
the absolute individual errors.
\subsection{Direct Summation}\label{sec:sum}
\ifpreprint\nobreak\noindent\fi
For small $N$ the exact force computation via direct summation is not only more
accurate than approximate methods but also more efficient. Therefore, we replace
the approximate technique by direct summation whenever the latter results in
higher accuracy at the same efficiency, see Appendix~B for more details.

If an interaction is executed by direct summation, the Taylor coefficients of
the interacting cell(s) are not affected, but the coefficients $\btC0$ and
$\btC1$ of the bodies within the cell(s) are accumulated.
\section{The Algorithm}\label{sec:alg}
\ifpreprint\vspace*{2mm}\fi
\subsection{Tree Building and Preparation}\label{sec:build}
\ifpreprint\noindent\fi
In the first stage, a hierarchical tree of cubic cells is build, as described in
\cite{BH}, albeit cells containing $s$ or less bodies are not divided. Next, the
tree is linked such that every cell holds the number of cell children%
\footnote{Hereafter `child' means a direct sub-node of a cell, while
  `descendant' refers to any node contained within a cell, including the
  children, the grand-children and so on.}, a pointer to its first cell child,
as well as the number of body children, of all body descendants, and a pointer
to the first body child. Since cell as well as body children are contiguous in
memory (to arrange this we actually use tiny copies of the bodies, called
`souls', that only hold a pointer to their body and its mass, position,
acceleration and potential) these data allow fast and easy access to all cell
and body children of any given cell and to all bodies contained in it.

Next, the masses $\sM{}$, centers of mass $\bz$, radii $r_{\rm max}$ and $r_{\rm
  crit}$, and specific quadrupole moments $\tM2$ of each cell are computed in a
recursive way from the properties of the child nodes.

\subsection{A new Generic Tree-Walk Algorithm}\label{sec:walk}
\ifpreprint\nobreak\noindent\fi
One important new feature of our code is the mutual treatment of all
interactions: both interacting nodes are source and sink simultaneously. The
standard tree-walk, as for instance implemented by the generic algorithm given
in \cite{WS}, as well as the the usual FMM coefficient accumulation algorithm,
e.g.\ in \cite{FMMb}, contain an inherent asymmetry between sinks and sources,
and thus cannot be used for our purposes. Instead, our algorithm approximates
the forces in two steps: an {\em interaction phase\/}, incorporating
\eqn{coeffs}, and an {\em evaluation phase\/}, incorporating \eqn{local-expn}.

\subsubsection{The Interaction Phase}\label{sec:iact}
\ifpreprint\nobreak\noindent\fi
Because of the mutuality of the interactions, we cannot accumulate the Taylor
coefficients $\btC{n}$ `on the walk', but each node must accumulate the
coefficients of all its interactions in its own private memory. Cells need
storage for $\btC0$ to $\btC3$, while bodies only need to accumulate $\btC0$ and
$\btC1$, i.e.\ potential and acceleration. The accumulation of these
coefficients is done by the following algorithm with the root cell for arguments
$A$ and $B$.
\begin{nonumalgorithm}[Interact(node $A$, node $B$)]
try to perform the interaction between{\sc node\/}s $A$ and $B$;%
\note{// see Appendix~B}
{\bf if}(it cannot be performed)
\   {\bf if} ($A=B$)%
\note{// split cell self-interaction}
\       {\bf for}(all pairs $\{a,b\}$ of child{\sc node\/}s of $A$)
\            {\sc Interact}($a,b$);
\   {\bf else if} ($r_{\rm max}(A) > r_{\rm max}(B)$)%
\note{// split bigger node}
\       {\bf for}(all child{\sc node\/}s $a$ of $A$)
\            {\sc Interact}($a,B$);
\   {\bf else}
\       {\bf for}(all child{\sc node\/}s $b$ of $B$)
\            {\sc Interact}($A,b$);
\end{nonumalgorithm}
Thus, if an interaction cannot be executed, using the formulae of the last
section -- see Appendix~B for details -- it is split.  In case of a mutual
interaction, the bigger node is divided and up to eight new mutual interactions
are created, while a self interaction of a cell results in up to 36 new
interactions between its child nodes. In practice, we use a non-recursive
code incorporating a stack of interactions.
\subsubsection{The Evaluation Phase}\label{sec:eval}
\ifpreprint\nobreak\noindent\fi
Finally, the Taylor coefficients relevant for each body are accumulated and the
expansion is evaluated at every body's position. After transforming $\btC{}$ to
$\bC{}$ for every cell and body, this is done by the following recursive
algorithm, which is initially called with the root cell and an empty Taylor
series as arguments.
\begin{nonumalgorithm}[Evaluate Gravity(cell $A$, Taylor series $T_0$)]
$T_A$ ={\sc Taylor series} due to the $\bC{n}$ of{\sc cell} $A$;
translate center of $T_0$ to center of mass of $A$;\note{// using \eqn{translate}}
$T_A$ += $T_0$;                                 \note{// add up coefficients}
{\bf for}(all{\sc body} children of $A$) \{
\    evaluate $T_A$ at{\sc body}'s position; \note{// as in \eqn{local-expn}}
\    add to{\sc body}'s potential and acceleration;
\}
{\bf for}(all{\sc cell} children $C$ of $A$)
\   {\sc Evaluate Gravity}($C$, $T_A$);         \note{// recursive call}
\end{nonumalgorithm}
Thus, the coefficients of the Taylor-series that is eventually evaluated at some
body's position have been added up from all hierarchies of the tree and hence
account for all interactions of all cells that contain the body.
\section{Error Assessment}\label{sec:assess}
\ifpreprint\nobreak\noindent\fi
Two types of force errors are involved in collisionless $N$-body simulations of
stellar dynamics. One is the unavoidable error between the smooth force field of
the underlying stellar system modeled and the forces {\em estimated\/} from the
positions of $N$ bodies (which are sampled from this stellar system). This {\em
  estimation error\/} can be reduced by increasing the number $N$ of bodies in
conjunction with a careful {\em softening\/}: the Newtonian Greens function
$g(\bx)=G/|\bx|$ is replaced with a non-singular function that approaches the
Newtonian form for $|\bx|$ larger than the softening length $\epsilon$. But at
fixed $N$, it cannot be decreased below a certain optimum value \cite{Merr,Db}.

The other type of error is introduced by an approximate rather than exact
computation of these estimated forces. While this {\em approximation error\/}
can be reduced to (almost) any size (at the price of increasing computational
effort), it is sufficient to reduce it well below the level of the estimation
error. We will now first assess the approximation error alone and then consider
the total error.
\begin{figure}
  \ifpreprint
    \centerline{\epsfxsize=11cm \epsfbox[40 171 574 697]{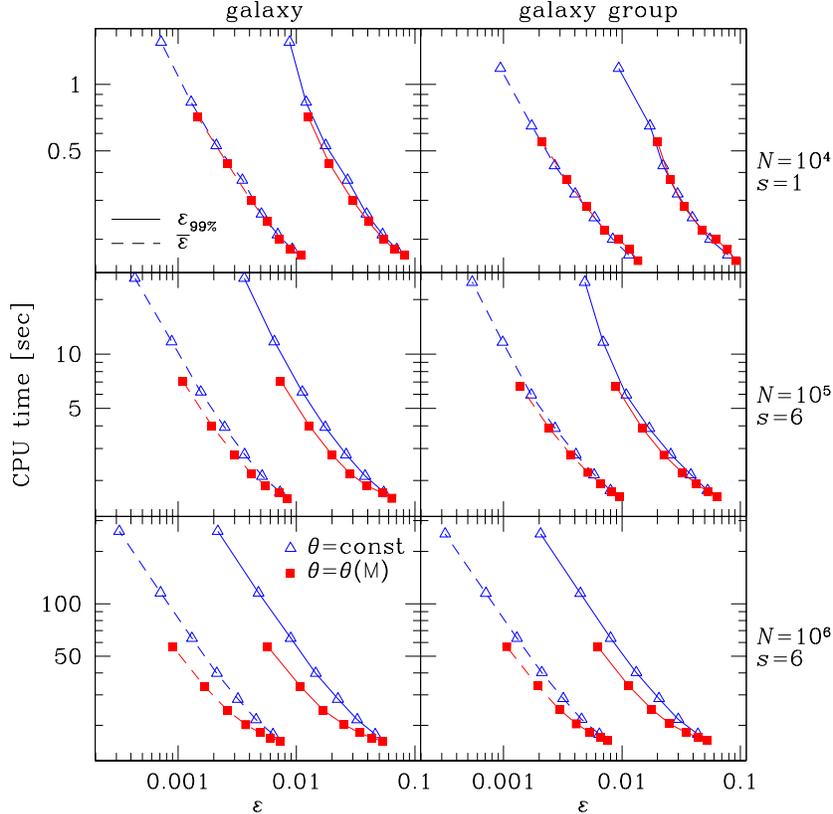}}
  \else
    \centerline{\epsfxsize=11cm \epsfbox[40 171 574 697]{Fig2.eps}}
  \fi
  \caption{The mean ({\em dashed}) and 99 percentile ({\em solid})
    relative force error versus the CPU time consumed on a PC (Pentium
    III/933MHz/Linux) for approximating the forces of a galaxy ({\em left}) and
    a group of galaxies ({\em right}), sampled with a total of $10^4$ ({\em
      top}), $10^5$ ({\em middle}), or $10^6$ ({\em bottom}) bodies. We used
    constant ({\em open triangles}) or mass-dependent tolerance parameter ({\em
      solid squares}). The symbols along each curve correspond, from left to
    right, to $\theta,\theta_{\rm min}=\,$0.3, 0.4, 0.5, 0.6, 0.7, 0.8, and 0.9.
    Cells containing $s$ or less bodies have not been divided.
    \label{fig:error}}
\end{figure}
\subsection{Accuracy of the Approximation}\label{sec:accur}
\ifpreprint\nobreak\noindent\fi
We estimate the accuracy of the approximated forces for various choices of the
parameters controlling the algorithm and for two typical astrophysical
situations: a spherical Hernquist \cite{Hern} model galaxy, which has
density and force per unit mass
\begin{equation} \label{hernquist}
  \rho(\bx) = {{\sf M}_{\rm tot} \over 2\pi}
  {r_0\over |\bx| \big(r_0+|\bx|\big)^3},\qquad
  \bF(\bx)  = - {\bx\over|\bx|}
  { G{\sf M}_{\rm tot}\over\big(r_0+|\bx|\big)^2},
\end{equation}
and a group of five such galaxies at different positions and with various scale
radii $r_0$. In either case, we sample $N=10^4$, $10^5$, and $10^6$ bodies. We
use standard Plummer softening, where $g(r)=G/\sqrt{r^2+\epsilon^2}$, with
softening lengths $\epsilon$ chosen such as to minimize the estimation error
\cite{Db}. In order to single out the approximation error, we compare with the
forces obtained by a computation via direct summation (in double precision; in
case of $N=10^6$ for the first $10^5$ bodies only). As measure for the relative
error, we compute for each body \cite{CM}
\begin{equation} \label{error}
  \varepsilon \equiv |a_{\rm approx} - a_{\rm direct}| / a_{\rm direct},
\end{equation}
where $a$ denotes the magnitude of the acceleration. \Fig{error} plots the mean
relative error, $\overline\varepsilon$, and that at the 99th percentile,
$\varepsilon_{99\%}$, versus the CPU time needed by an ordinary PC (Pentium
III/933MHz/Linux/compiler: gcc version 2.95.2) for both constant and
mass-dependent tolerance parameter. This figure allows several interesting
observations.
\begin{enumerate}
\item For the same $\theta=\,$const and stellar system, the errors decrease with
  increasing $N$. This is because, at constant {\em relative\/} force error per
  individual interaction (as is the case for $\theta=\,$const), the total error
  of some body's force scales with the inverse square root of the number of
  individual interactions contributing, which increases with $N$.
\item At the same operational effort, as measured by the CPU time, the
  mass-dependent tolerance parameter employing \eqn{open:mass} results in
  smaller errors than $\theta=\,$const, for $N>10^4$. This advantage becomes
  more pronounced for larger $N$, because, with $\theta=\,$const, the {\em
    absolute\/} force errors of individual interactions contributing to some
  body's force vary stronger with increasing $N$ (due to the larger range of
  cell masses), such that balancing them becomes more beneficial.
\item Finally, a relative error of $\varepsilon_{99\%}$ of a few per cent or
  $\overline\varepsilon$ of a few $10^{-3}$ at $N=10^5$, which is commonly
  accepted to be sufficient in astrophysical contexts \cite{CM}, requires a
  tolerance parameter $\theta=\,$const$\,\simeq0.65$ or $\theta_{\rm
    min}\simeq0.5$.
\end{enumerate}
\subsection{The Total Force Error}\label{sec:error}
\ifpreprint\nobreak\noindent\fi
The important question here is for which choices of $\theta$ is the
approximation error negligible compared to the estimation error? To answer this
question, we have performed some experiments using samples of $N=10^4$, $10^5$
and $10^6$ bodies drawn from a Hernquist model and computed the mean averaged
squared error (MASE) of the force (per unit mass) \cite{AT}:
\begin{equation} \label{mase}
  {\rm MASE}(\bF) = \left\{{1\over{\sf M_{\rm tot}}} \sum_i \mu_i
    \big(\hat{\bF}_i - \bF(\bx_i)\big)^2 \right\}.
\end{equation}
Here, $\hat{\bF}_i$ and $\bF(\bx)$ are, respectively, the approximately
estimated force for the $i$th body and the true force field of the stellar
system. The curly brackets denote the ensemble average over many possible random
realizations of the same underlying stellar density by $N$ bodies. We used
$10^7/N$ ensembles and computed the MASE$(\bF)$\footnote{The force field of the
  Hernquist model has a central singularity, causing a 100\% force error at
  $r=0$, which cannot be resolved by $N$-body methods. In our experiments, we
  have therefore restricted the summation in \eqn{mase} to
  $|\bx_i|>\epsilon/2$.}  for various values of $\theta$, but always at optimum
$\epsilon$ \cite{Db}. As one might already have guessed from the behavior of the
approximation error in \Fig{error}, the relative increase of the MASE$(\bF)$ is
negligible: even for $\theta=0.9$, the approximation error contributes less than
one per cent, in agreement with the findings of \cite{AT}. Based on this result,
one may advocate the usage of tolerance parameters larger than $\theta_{\rm
  min}\simeq0.5$.  However, the distribution of approximation errors is not
normal, and the rms error, which is essentially measured by (the square root of)
the MASE$(\bF)$, may well underestimate the danger of using large $\theta$. We
therefore cannot recommend using $\theta_{\rm min}\ga0.7$.
\section{Performance Tests}\label{sec:perform}
\ifpreprint\vspace*{2mm}\fi
\subsection{Scaling with \boldmath $N$}\label{sec:scaling}
\begin{figure}
  \ifpreprint
    \centerline{\epsfxsize=85mm \epsfbox[24 307 581 706]{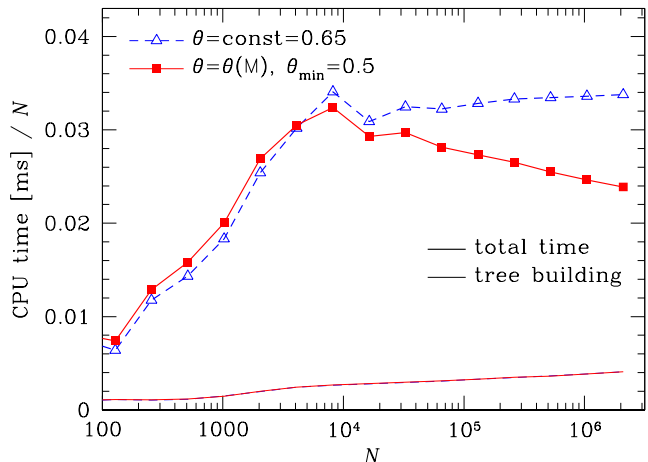}}
  \else
    \centerline{\epsfxsize=85mm \epsfbox[24 307 581 706]{Fig3.eps}}
  \fi
  \caption{CPU time consumption per body (Pentium III/933MHz PC) versus
    $N$ for the test case of a galaxy group, $s=6$, and
    $\theta=\,$const$\,=0.65$ or $\theta_{\rm min}=0.5$. \label{fig:N}}
\end{figure}

\ifpreprint\nobreak\noindent\fi
We measured the CPU time consumption for both constant $\theta=0.65$ and
$\theta=\theta(\sM{})$ with $\theta_{\rm min}=0.5$, using $s=6$ in either case.
\Fig{N} plots the consumed time (averaged over many experiments) per body versus
$N$ plotted on a logarithmic scale. For the case of $\theta=\theta(\sM{})$,
\Tab{N} gives the number of cells as well as the number of individual body-body
(B-B), cell-body (C-B) and cell-cell (C-C) interactions, where the latter two
are split into those done via a Taylor expansion and direct summation
(subscripts `app' and `dir'), respectively.

\begin{table}
  \caption{Number of Cells and of Interactions for $\theta_{\rm min}=0.5$,
    $s=6$ and the same test case as in \Fig{N}\label{tab:N}}
 \begin{tabular*}{\textwidth}{@{\extracolsep{\fill}}rrrrrrrr}
 \hline
   $N$ & $N_{\rm cells}$ & B-B &
   C-B$_{\rm app}$ & C-B$_{\rm dir}$ &
   C-C$_{\rm app}$ & C-C$_{\rm dir}$ & total \cr
 \hline
    1000 &     370 & 210 &    3746 &    2049 &     4057 &    1600 &    11662 \cr
    3000 &    1090 & 472 &   16870 &    5184 &    30101 &    4472 &    57099 \cr
   10000 &    3558 & 753 &   61873 &   11914 &   163554 &   11887 &   249981 \cr
   30000 &   10607 & 494 &  170634 &   24984 &   572322 &   20102 &   790474 \cr
  100000 &   35282 & 112 &  429035 &   73350 &  1954508 &   53466 &  2516146 \cr
  300000 &  106065 &  66 & 1041836 &  205821 &  5457445 &  137138 &  6859114 \cr
 1000000 &  353342 &   1 & 2918326 &  621802 & 16105065 &  393311 & 19023391 \cr
 3000000 & 1060650 &   2 & 7771584 & 1689115 & 42974890 & 1047627 & 53645379 \cr
 \hline
 \end{tabular*}
\end{table}

The tree code requires ${\cal O}(N\log N)$ operations, corresponding to a rising
straight line in \fig{N}. For our code, however, there is a turn-over at
$N\sim10^4$, above which the CPU time per body approaches a constant\footnote{
  The slow rise can be entirely attributed to the tree-building, which is an
  ${\cal O}(N\log N)$ process.} (for $\theta=\,$const) or even decreases with
$N$ (for $\theta=\theta(\sM{})$), i.e.\ the total number of operations becomes
${\cal O}(N)$ or less, which is also evident from the number of interactions in
\Tab{N}.

In order to understand these scalings, let us first consider the simpler case of
$\theta=\,$const and a homogeneous distribution of bodies. Then, eight-folding
$N$ is equivalent to arranging eight copies of the old root cell into the
octants of the new root cell \cite{BH}, and the total number of interactions
rises from $N_I$ to $8N_I+N_+$, where $N_+$ interactions are needed for the
mutual forces between these octants. In terms of a differential equation, this
gives
\begin{equation} \label{ode}
        {{\rm d} N_I\over{\rm d} N}
        \simeq  {N_I\over N} {\Delta\ln N_I\over\Delta\ln N}
        \approx {N_I\over N} + {N_+\over N8\ln 8},
\end{equation}
where the first term on the right-hand side accounts for the {\em
  intra-domain\/} and the second for the {\em inter-domain\/} interactions.
\Eqn{ode} has solution
\begin{equation} \label{sol}
        N_I \simeq c_0 N + {N\over8\ln8} \int {N_+\over N^2}\;{\rm d} N.
\end{equation}
In the tree code, every body requires a constant number of additional
interactions, i.e.\ $N_+\propto N$, and the second term in \eqb{sol} becomes
$\propto N\ln N$ dominating $N_I(N)$. However, in the new code, $N_+$ grows
sub-linear for large $N$, since a constant number of cell-cell interactions
accounts for most new interactions of all bodies. In this case, the second term
on the right-hand side of \eqn{sol} also grows sub-linear with $N$ and $N_I$
will eventually be dominated by the first term. That is, in contrast to the tree
code, the inter-domain interactions are neglible at large $N$ when compared to
the intra-domain interactions. The transition value of $N$ will depend on the
tolerance parameter $\theta$ and the distribution of bodies.

Another way of estimating the scaling of the computational costs with $N$ is
similar to the FMM approach \cite{FMMa,FMMb}: On each level $l$ of the tree
there are $\sim8^l$ cells of mass $\sM{}\sim8^{-l}$, i.e.\ $n(\sM{})\propto{\sf
  M}^{-2}$. Each cell has $\propto\theta^{-3}$ interactions, and thus
\begin{equation} \label{scaling}
        N_I \propto \int { d\sM{}\over{\sf M}^{2}\theta^3}.
\end{equation}
Hence, for $\theta=\,$const, $N_I\propto1/{\sf M}_{\rm min}\propto N$, while a
shallower scaling results if $\theta(\sM{})$ increases towards smaller masses.
Empirically we find for $N>30000$ that the CPU time used by the force
computation alone (without tree building) is very well fit by the power-law
$\propto N^{0.929\pm0.001}$.
\begin{figure}
  \ifpreprint
    \centerline{\epsfxsize=90mm \epsfbox[29 309 576 706]{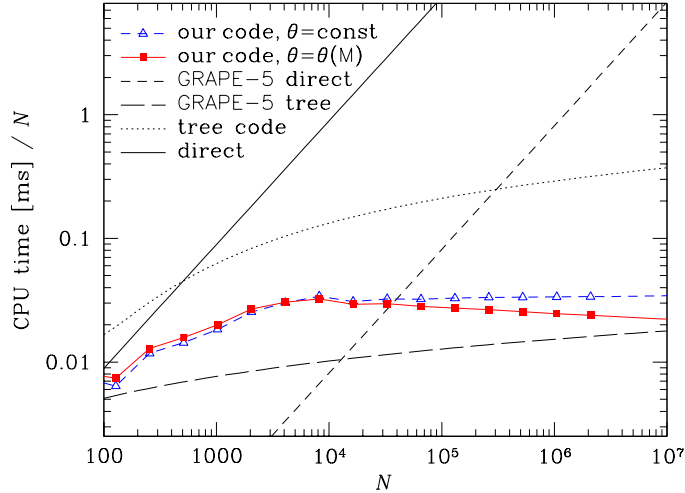}}
  \else
    \centerline{\epsfxsize=90mm \epsfbox[29 309 576 706]{Fig4.eps}}
  \fi
  \caption{CPU time per body versus $N$ for various techniques. The
    direct, tree and our code were all compiled and run with the same
    equipment. For the new code, we use $\theta=0.65$ and $\theta_{\rm
      min} =0.5$, while for the tree code $\theta=0.8$ and $p=3$.  The
    scalings for the {\sf GRAPE-5} methods are taken from \cite{GRAPE}.
    The accuracy requirements for the approximate (non-direct) methods are
    adapted for astrophysical applications.
    \label{fig:var}}
\end{figure}
\subsection{Comparison with Other Methods used in Stellar Dynamics}
\label{sec:comp}
\ifpreprint\nobreak\noindent\fi
For various computational techniques used in stellar dynamics, \Fig{var} plots
$N$ versus the CPU time consumption normalized by $N$, both on logarithmic
scales. An ordinary direct method running on a general-purpose computer is
slower than our code for any $N\ga100$. The {\sf GRAPE-5} system \cite{GRAPE}
obtains a $\sim100$ times higher performance by wiring elementary gravity into
special-purpose hardware and, for $N\la10^4$, is faster than any other method.

The new code presented here is the fastest method running solely on
general-purpose computers and the only one faster than ${\cal O}(N\log N)$. In
particular, it out-performs the popular tree code by a factor of 10 and more.
At $30000\la N\la3\times10^7$, the only technique that requires less CPU time is a
combination of the tree code with a {\sf GRAPE-5} board \cite{GRAPE,G-tree}.
Here, the speed-up due to the usage of special-purpose hardware does not quite
reach that for direct methods, because of tree-building and other overheads that
cannot be done on the {\sf GRAPE} board.
\subsection{Comparison with Fast Multipole Methods}
\label{sec:fmm}
\ifpreprint\nobreak\noindent\fi
Because our code relies on cell-cell interactions, it may be considered a
variant of FMM, introduced by Greengard \& Rokhlin \cite{FMMa,FMMa1}. However,
it differs in several ways from most implementations of FMM. (i) the expansions
are centered on the cells' centers of mass instead of the geometrical centers;
(ii) a Cartesian Taylor series is used instead of an expansion in spherical
harmonics; (iii) the interaction partners are determined by a multipole acceptance
criterion rather than by their mutual grid position; (iv) the mutuality of
interactions is fully exploited; (v) the expansion order $p$ is fixed and the
accuracy controlled by the parameter $\theta$.

\begin{table}
  \caption{Comparison with FMM: Timing Results for Bodies Uniformly
    Distributed in a Cube\label{tab:compare}}
 \begin{tabular*}{\textwidth}{@{\extracolsep{\fill}}rrrlrrl}
 \hline
   \multicolumn{1}{c}{$N$} &
   \multicolumn{1}{c}{$T_{\rm FMM}^a$} &
   \multicolumn{1}{c}{$T_{\rm direct}^a$} &
   \multicolumn{1}{c}{$E^a$} &
   \multicolumn{1}{c}{$T_{\rm approx}^b$} &
   \multicolumn{1}{c}{$T_{\rm direct}^c$} &
   \multicolumn{1}{c}{$E^b$} \cr
 \hline
   20000 &13.3            &   233 &$7.9\cdot10^{-4}$
   & 0.97 &   136 &$3.7\cdot10^{-4}$\cr
   50000 &27.7            &  1483 &$5.2\cdot10^{-4}$
   & 2.64 &   924 &$3.3\cdot10^{-4}$\cr
  200000 &158\phantom{.0} & 24330 &$8.4\cdot10^{-4}$
   &10.77 & 14694 &$3.4\cdot10^{-4}$\cr
  500000 &268\phantom{.0} &138380 &$7.0\cdot10^{-4}$
   &29.42 & 91134 &$3.7\cdot10^{-4}$\cr
 1000000 &655\phantom{.0} &563900 &$7.1\cdot10^{-4}$
   &58.34 &366218 &$3.5\cdot10^{-4}$\cr
 \hline
 \end{tabular*}
 \tablenotes{
   $^a$ Using FMM; data from Table~I of Cheng et al.\ \cite{FMMb}\par
   $^b$ Using the code presented here on a computer identical to that used by
   Cheng et al.\par
   $^c$ Using our own implementation of direct summation on the same computer}
\end{table}
To assess the effect of these differences, we compared our code directly with
the 3D adaptive FMM code by Cheng, Greengard \& Rokhlin \cite{FMMb}. We
performed a test identical to one reported by these authors ($N$ bodies randomly
distributed in a cube) on an identical computer (a Sun UltraSPARC with 167MHz)
using the same error measure (eq.~(57) of \cite{FMMb}):
\begin{equation}
  E = \left[\sum_i(\Phi_i - \tilde\Phi_i)^2 \Big/ \sum_i \Phi_i^2\right]^{1/2}
\end{equation}
where $\Phi$ and $\tilde\Phi$ are the potential computed by direct summation and
the approximate method, respectively (unfortunately, Cheng et~al.\ do not give
the more relevant error of the accelerations). \Tab{compare} gives, in the last
three columns, the CPU time ($T_{\rm approx}$) in seconds and error $E$ for our
code with $\theta=1$ and $s=6$ as well as the time needed for direct summation
in 64bit precision ($T_{\rm direct}$); columns 2-4 report the data from Table~I
of \cite{FMMb}. On average our code is faster by more than a factor of ten and
twice as accurate (even though we compromised the approximation of the potential
by omitting the octopole contributions).

What causes this enormous difference? Since for the direct summation the timings
are much more similar, we can rule out differences in hardware, compiler, etc.\ 
as cause. An important clue is the fact that our code cannot compete with the
Cheng et al.\ FMM in the regime $E\la10^{-6}$. While our code tries to reach
this goal by decreasing $\theta$, FMM obtains it by increasing $p$. In general,
the accuracy as well as the performance are controlled by both the order $p$ and
the choice of interaction partners, parameterized in our code by $\theta$.
Hence, maximal efficiency at given accuracy is obtained at a unique choice of
$(p,\theta)$. Apparently, low orders $p$ are optimal for low accuracies,
while high accuracies are most efficiently obtained with high orders, instead
of an increasing number of interactions (decreasing $\theta$).

Thus, traditional FMM is less useful in the low-accuracy regime, such as needed
in stellar dynamics, in agreement with earlier findings \cite{CM}, and our code
may be called a variant of FMM optimized for low-accuracy. Clearly, however, a
code for which $p$ and $\theta$ can be adapted simultaneously would be superior
to both.
\section{Discussion and Summary} \label{sec:summ}
\ifpreprint\nobreak\noindent\fi
Our code for the approximate computation of mutual long-range forces between $N$
bodies extends the traditional Barnes \& Hut \cite{BH} tree code by including
cell-cell interactions, similar to fast multipole methods (FMM). However, unlike
most implementations of FMM, our code is optimized for comparably low-accuracy,
which is sufficient in stellar dynamical applications (\Sec{fmm}).

As a unique feature, our code exploits the mutual character of gravity: both
nodes of any interaction are sink and source simultaneously. This results in
exact conservation of Newton's third law and substantially reduces the
computational effort, but requires a novel tree-walking algorithm (\Sec{walk})
which preserves the natural symmetry of each interaction. Note that the generic
algorithm given in \Sec{walk} is not restricted to long-range force
approximations, but can be used for any task that incorporates mutuality, for
instance neighbor and collision-partner searching.
\subsection{Complexity}\label{sec:scal}
\ifpreprint\nobreak\noindent\fi
The new code requires ${\cal O}(N)$ or less operations (\fig{N}) for the
approximate computation of the forces of $N$ mutually gravitating bodies. For
stellar dynamicists, this is the first competitive code better than ${\cal
  O}(N\log N)$. A complexity of ${\cal O}(N)$ was expected for methods based on
cell sinks (implying cell-cell interactions) \cite{FMMa,FMMb,WS}, but, to my
knowledge, hardly ever shown empirically in three dimensions.

Our code obtains a complexity of less than ${\cal O}(N)$ by employing a
mass-dependent tolerance parameter $\theta$. The traditional $\theta=\,$constant
results in equal {\em relative\/} errors of each interaction, such that the
total error of any body's force is dominated by the interactions with the most
massive cells. By slightly increasing the tolerance parameter for less massive
cells, we obtain (approximately) equal {\em absolute\/} errors, resulting in a
lower total force error than the traditional method for the same number of
interactions. Thus, at the same error, we require less interactions.  The
additional interactions, arising when increasing $N$, occur at ever less massive
cells and hence at ever larger tolerance parameters. This causes the computational
costs to rise sub-linear with $N\ga10^4$ (depending on the accuracy
requirements).
\subsection{Performance}\label{sec:per}
\ifpreprint\nobreak\noindent\fi
We have shown that on general-purpose computers our code out-performs any
competitor code commonly used in the field of stellar dynamics. A recent
adaptive 3D implementation of FMM \cite{FMMb} is also out-performed by a factor
of ten (\Sec{fmm}), which is related to the fact that traditional FMM codes
appear to be good only in the high-accuracy regime. The code presented here was
optimized for low accuracies, but by increasing the expansion order $p$, one can
easily obtain a version suitable for high accuracies, and it remains to be seen
how it would perform compared to traditional FMM.

Currently, the only faster method appears to be a {\sf GRAPE}-supported tree
code \cite{G-tree}, which uses the special-purpose {\sf GRAPE} hardware
\cite{GRAPE}. Unfortunately, unlike the tree code, our code cannot be combined
with the current {\sf GRAPE} hardware. There are, however, no conceptual
obstacles against hard-wiring \eqns{hard-coeffs} into special-purpose hardware,
which should yield a speed-up comparable to that of tree to {\sf GRAPE} tree,
i.e.\ a factor $\sim50$.
\subsection{Publication of the Code}\label{sec:pub}
\ifpreprint\nobreak\noindent\fi
Our code is written in {\sf C++}, also includes a purely two-dimensional version
(not described here), and is linkable to {\sf C} and {\tt FORTRAN} programs. The
code \ifpreprint will be \else is \fi publicly available from the author upon
request.

A full $N$-body code based on this force algorithm is available under the {\sc
  NEMO\/} \cite{nemo} package (http://bima.astro.umd.edu/nemo).
\appendix{a}
\ifpreprint\nobreak\noindent\fi
Here, we give an error estimate for the Taylor series approximation of
gravity. Using the integral form of the Taylor series remainder, we find for the
remainder in \eqn{Taylor}, with $\B\Delta\equiv(\bx-\bz_\rA)-(\by-\bz_\rB)=
\bx-\by-\bR$,
\begin{eqnarray} \label{remain:pot}
  {\cal R}_p(g) &=&  {\B\Delta^{(p+1)}\over p!}\,\odot
  \int_0^1 dt\,(1-t)^p\,\B\nabla^{(p+1)} g(\bR+\B\Delta t)
\\[1ex] \label{remain:acc}
  \B\nabla{\cal R}_p(g) &=& {\B\Delta^{(p)}\over(p-1)!}\,\odot
  \int_0^1 dt\,(1-t)^{p-1}\,\B\nabla^{(p+1)} g(\bR+\B\Delta t),
\end{eqnarray}
For Newtonian gravity,
\begin{eqnarray}
  \left\| {1\over p!}\int_0^1dt\,(1-t)^p\,
    \B\nabla^{(p+1)}g(\bR+\B\Delta t) \right\|
  &\le&   {1\over(R-|\B\Delta|)\,R^{p+1}},          \\[1ex]
  \left\| {1\over (p-1)!}\int_0^1dt\,(1-t)^{p-1}\,
    \B\nabla^{(p+1)}g(\bR+\B\Delta t) \right\|
  &\le&   {(p+1)R - p|\B\Delta| \over (R-|\B\Delta|)^2\,R^{p+1}}.
\end{eqnarray}
For the summation over the source cell, we find
\begin{equation}
  \Big\| \sum_{\Bs{y}_i\in\rB} \mu_i \B\Delta^{(p)} \Big\|
  \le \sum_{k=0}^p {p\choose k} r_{\rm A,max}^k\,\big\|\bM{(p-k)}_\rB\big\|
  \le (r_{\rm A,max}+r_{\rm B,max})^p \sM{}_\rB
\end{equation}
with $\sM{}_\rB$ the mass of cell B. With $\theta>(r_{\rm A,max}+r_{\rm
  B,max})/R$, we finally get
\begin{eqnarray} \label{error:pot}
  |{\cal R}_p(\Phi_{\rB\to\rA})| &\le&
  {\theta^{p+1}\over1-\theta} {\sM{}_\rB\over R} \\ \label{error:acc}
  |{\cal R}_p(\B\nabla\Phi_{\rB\to\rA})| &\le&
  {(p+1)\theta^p\over(1-\theta)^2}        {\sM{}_\rB\over R^2}.
\end{eqnarray}

\appendix{b}
\ifpreprint\nobreak\noindent\fi
Here, we give the interaction details for {\sc Algorithm 1} in \Sec{iact}.
Mutual body-body interactions are done by elementary gravity, while body
self-interactions are ignored. Mutual interactions between nodes containing
$N_1$ and $N_2$ bodies are treated as follows.
\begin{enumerate}
\item If $N_1 N_2 < N_{\rm nn}^{\rm pre}$, execute the interaction by direct
  summation; otherwise,
\item if the interaction is well-separated, execute it using \eqns{hard-coeffs};
  otherwise,
\item if $N_1 N_2 < N_{\rm nn}^{\rm post}$, execute the interaction by direct
  summation; otherwise,
\item the interaction cannot be executed, but must be split.
\end{enumerate}
Here, $N_{\rm nn}^{\rm pre}$ and $N_{\rm nn}^{\rm post}$ have different values
depending whether it is a cell-body (cb) or cell-cell (cc) interaction (see
below). Cell self-interactions are done slightly differently:
\begin{enumerate}
\item If $N_1 < N_{\rm cs}$, execute the interaction by direct summation;
  otherwise,
\item the interaction cannot be executed, but must be split.
\end{enumerate}
The numbers $N_{\rm cb}^{\rm pre}$, $N_{\rm cb}^{\rm post}$, $N_{\rm cc}^{\rm
  pre}$, $N_{\rm cc}^{\rm post}$, and $N_{\rm cs}$ determine the usage of direct
summation. After some experiments, I found the following values to result in
most efficient code at given accuracy (but this certainly depends on the
implementation details).
\begin{equation}
  \begin{array}{lcrlcrlcr}
    N_{\rm cb}^{\rm pre}  &=&  3,& 
    N_{\rm cc}^{\rm pre}  &=&  0,& \\
    N_{\rm cb}^{\rm post} &=&128,&
    N_{\rm cc}^{\rm post} &=& 16,&
    N_{\rm cs}            &=& 64.
  \end{array}
\end{equation}
Note that cell-body interactions with as many as 128 bodies in the cell hardly
ever occur, since the interaction algorithm favors interactions between roughly
equally sized nodes. Thus, cell-body interactions will almost always be
executed.
\begin{acknowledgments}
  I thank J.~Makino from Tokoy University for helpful discussions and R.~Ibata
  and C.~Pichon from Strasbourg Observatory for hardware support.
\end{acknowledgments}

\end{document}